\documentstyle [12pt] {article}

\catcode`@=12
\begin{document}
\thispagestyle{empty}
\begin{flushright} UCRHEP-T244\\February 1999\
\end{flushright}
\vspace{1.0in}
\begin{center}
{\Large \bf Supersymmetry and Neutrino Masses\\}
\vspace{1.5in}
{\bf Ernest Ma\\}
\vspace{0.3in}
{\sl Physics Department, University of California,\\ 
Riverside, CA 92521, USA\\}
\vspace{1.5in}
\end{center}
\begin{abstract}\
Neutrino masses are zero in the minimal supersymmetric standard model.  
I discuss how they may become nonzero with new interactions which may or 
may not violate $R$-parity conservation.
\end{abstract}
\vspace{0.5in}
-----------------------\\
\noindent Talk presented at the 6th Corfu Summer Institute on Elementary 
Particle Physics, 1998.

\newpage
\baselineskip 24pt

\section{Introduction}

On the strength of the recent report of atmospheric neutrino oscillations 
\cite{1}, as well as previous other indications of solar \cite{2} and 
accelerator \cite{3} neutrino oscillations, neutrino masses are now 
considered to be almost established experimentally.  Yet there is no 
clear theoretical consensus as to the origin of neutrino masses.  In the 
standard model, the usual way is to add three right-handed neutrino singlets 
with large Majorana masses and use the canonical seesaw mechanism \cite{4} 
to obtain small Majorana masses for $\nu_e$, $\nu_\mu$, and $\nu_\tau$. 
On the other hand, other mechanisms are available \cite{5}, the simplest 
alternative being the addition of a heavy scalar triplet \cite{6}.

There is another important theoretical reason for going beyond the minimal 
standard model, {\it i.e.} supersymmetry.  However, the minimal supersymmetric 
standard model (MSSM) keeps the neutrinos massless because it contains no 
extra fields or interactions which could make them massive.  Of course, one 
may simply add \cite{7} three right-handed neutrino singlet superfields to 
the MSSM and invoke the canonical seesaw mechanism as before.  On the other 
hand, given the particle content of the MSSM, one may also allow new, 
lepton-number nonconserving terms in the Lagrangian which would then induce 
nonzero neutrino masses \cite{8,9}.  In this talk, I will review briefly 
this latter situation where $R$-parity is usually assumed to be violated, and 
point out its potential problem with leptogenesis, ending with a proposal 
of radiative neutrino masses with $R$-parity conservation. 

\section{MSSM and $R$-Parity}

The well-known superfield content of the MSSM is given by
\begin{eqnarray}
&& Q_i = (u_i, d_i)_L \sim (3,2,1/6), \\ 
&& u^c_i \sim (3^*,1,-2/3), \\ 
&& d^c_i \sim (3^*,1,1/3), \\ 
&& L_i = (\nu_i, l_i)_L \sim (1,2,-1/2), \\ 
&& l^c_i \sim (1,1,1); \\ 
&& H_1 = (\bar \phi^0_1, -\phi^-_1) \sim (1,2,-1/2), \\ 
&& H_2 = (\phi^+_2, \phi^0_2) \sim (1,2,1/2).
\end{eqnarray}
Given the above transformations under the standard $SU(3) \times SU(2) \times 
U(1)$ gauge group, the corresponding superpotential should contain in general 
all gauge-invariant bilinear and trilinear combinations of the superfields. 
However, to forbid the nonconservation of both baryon number ($B$) and lepton 
number ($L$), each particle is usually assigned a dicrete $R$-parity:
\begin{equation}
R \equiv (-1)^{3B+L+2j},
\end{equation}
which is assumed to be conserved by the allowed interactions.  Hence the 
MSSM superpotential has only the terms $H_1 H_2$, $H_1 L_i l^c_j$, 
$H_1 Q_i d^c_j$, and $H_2 Q_i u^c_j$.  Since the superfield $\nu^c_i \sim 
(1,1,0)$ is absent, $m_\nu = 0$ in the MSSM as in the minimal standard model. 
Neutrino oscillations \cite{1,2,3} are thus unexplained.

Phenomenologically, it makes sense to require only $B$ conservation (to make 
sure that the proton is stable), but to allow $L$ violation (hence $R$-parity 
violation) so that the additional terms $L_i H_2$, $L_i L_j l^c_k$, and 
$L_i Q_j d^c_k$ may occur.  Note that they all have $\Delta L = 1$. 
From the bilinear terms \cite{9}
\begin{equation}
-\mu H_1 H_2 + \epsilon_i L_i H_2,
\end{equation}
we get a $7 \times 7$ neutralino-neutrino mass matrix
\begin{equation}
\left[ \begin{array}{c@{\quad}c@{\quad}c@{\quad}c@{\quad}c} M_1 & 0 & 
-g_1 v_1 & -g_1 v_2 & -g_1 u_i \\ 0 & M_2 & g_2 v_1 & -g_2 v_2 & 
g_2 u_i \\ -g_1 v_1 & g_2 v_1 & 0 & -\mu & 0 \\ g_1 v_2 & -g_2 v_2 
& -\mu & 0 & \epsilon_i \\ -g_1 u_i & g_2 u_i & 0 & \epsilon_i & 0 
\end{array} \right],
\end{equation}
where $v_{1,2} = \langle \phi^0_{1,2} \rangle /2$ and $u_i = \langle \tilde 
\nu_i \rangle /2$, with $i = e, \mu, \tau$.  Note first the important fact 
that a nonzero $\epsilon_i$ implies a nonzero $u_i$ \cite{9}. Note also that 
if $u_i/\epsilon_i$ is the same for all $i$, then only one linear combination 
of the three neutrinos gets a tree-level mass.  From the trilinear terms, 
neutrino masses are also obtained \cite{8,10}, now as one-loop radiative 
corrections.  Note that these occur as the result of supersymmetry breaking 
and are suppressed by $m_d^2$ or $m_l^2$.

\section{$L$ Nonconservation and the Universe}

As noted earlier, the $R$-parity nonconserving interactions have $\Delta L 
= 1$.  Furthermore, the particles involved have masses at most equal to the 
supersymmetry breaking scale, {\it i.e.}  a few TeV.  This means that their 
$L$ violation together with the $B + L$ violation by sphalerons \cite{11} 
would erase any primordial $B$ or $L$ asymmetry of the Universe \cite{12,13}. 
To avoid such a possibility, one may reduce the relevant Yukawa couplings 
to less than about $10^{-7}$, but a typical minimum value of $10^{-4}$ is 
required for realistic neutrino masses.  Hence the existence of the present 
baryon asymmetry of the Universe is unexplained if neutrino masses originate 
from these $\Delta L = 1$ interactions.  This is a generic problem of all 
models of radiative neutrino masses where the $L$ violation can be traced 
to interactions occuring at energies below $10^{13}$ GeV or so.

Consider the prototype (Zee) model of radiative neutrino masses \cite{14}.  
It is not supersymmetric and it only adds one charged scalar singlet 
$\chi^\pm$ and a second Higgs doublet to the minimal standard model.  Call 
the two Higgs doublets $\Phi_{1,2}$, then the trilinear coupling $\chi^- 
(\phi^+_1 \phi^0_2 - \phi^0_1 \phi^+_2)$ is allowed as well as the Yukawa 
coupling $\chi^+ (\nu_i l_j - l_i \nu_j)$.  Hence there is an effective 
dimension-5 operator $\nu_i \nu_j \phi^0_1 \phi^0_2$ which renders the 
neutrinos massive \cite{5}, but it is again suppressed by $m_l^2$.  Note 
that the new interactions have $\Delta L = 2$.

\section{Supersymmetric Radiative Neutrino Masses and\\ Leptogenesis}

It has been shown recently \cite{6} that naturally small Majorana neutrino 
masses may be obtained from heavy scalar triplets and if the latter have 
masses of order $10^{13}$ GeV, their decays could generate a lepton 
asymmetry which then gets converted into the present baryon asymmetry of 
the Universe through the electroweak phase transition.  The same role 
may be attributed to the scalar singlets of the Zee model if 
they are heavy enough, but then to obtain realistic radiative neutrino 
masses, unsuppressed Yukawa couplings are needed.

Consider now the following supersymmetric extension of the Zee model.  Since 
all the interactions are either $\Delta L = 0$ or $\Delta L = 2$, $R$-parity 
is conserved.  Because of the requirement of supersymmetry, there is a 
doubling of the scalar superfields:
\begin{eqnarray}
&& \chi_1^+ \sim (1,1,1;+,-), \\
&& \chi_2^- \sim (1,1,-1;+,-), \\ 
&& H_{1,3} \sim (1,2,-1/2;+,\pm), \\ 
&& H_{2,4} \sim (1,2,1/2;+,\pm).
\end{eqnarray}
A fourth family of leptons is then added:
\begin{eqnarray}
&& (N_1^0,E^-) \sim (1,2,-1/2;-,-), \\ 
&& N_2^0 \sim (1,1,0;-,-), \\
&& E^+ \sim (1,1,1;-,-).
\end{eqnarray}
In the above, the assignments of these superfields under a discrete 
$Z_2 \times Z'_2$ symmetry are also displayed.  The first is merely the 
one usually assumed to obtain $R$-parity; the second is used to distinguish 
the new particles from those of the MSSM.  The relevant terms in the 
$R$-parity preserving superpotential of this model are then given by
\begin{eqnarray}
W &=& \mu_{12} (h_1^0 h_2^0 - h_1^- h_2^+) \nonumber \\ 
&+& \mu_{34} (h_3^0 h_4^0 - h_3^- h_4^+) \nonumber \\ 
&+& m_\chi \chi^+ \chi^- \nonumber \\ 
&+& (m_E/v_1)(h_1^0 E^- - h^-_1 N_1^0) E^+ \nonumber \\ 
&+& f_i (\nu_i h_3^- - l_i h_3^0) E^+ \nonumber \\ 
&+& f'_j (\nu_j E^- - l_j N_1^0) \chi^+_1 \nonumber \\ 
&+& f_{24} (h_2^+ h_4^0 - h_2^0 h_4^+) \chi_2^-,
\end{eqnarray}
where $v_{1,2}$ are the vacuum expectation values of $h_{1,2}^0$.  The 
unsuppressed one-loop diagram generating neutrino masses is shown in 
Fig.~1 of Ref.~[13].  Note that the effective supersymmetric dimension-5 
operator $L_i L_j H_2 H_2$ is indeed realized.  Assuming that the masses 
of the scalar leptons of the fourth family to be equal to $M_{SUSY}$, the 
neutrino mass matrix is then obtained:
\begin{equation}
{(f_i f'_j + f'_i f_j) f_{24} v_2^2 m_E \mu_{12} \mu_{34} \over 16 \pi^2 v_1 
M_{SUSY}^2 m_\chi} \ln {m_\chi^2 \over M_{SUSY}^2}.
\end{equation}
To get an estimate of the above expression, let $f_i = f'_j = f_{24} = 1$, 
$m_E = v_1$, $\mu_{12} = \mu_{34} = M_{SUSY}$, then
\begin{equation}
m_\nu =  {1 \over 8 \pi^2} {v_2^2 \over m_\chi} \ln {m_\chi^2 \over 
M_{SUSY}^2}.
\end{equation}
Assuming $v_2 \sim 10^2$ GeV, $m_\chi \sim 10^{13}$ GeV, and $M_{SUSY} 
\sim 10^3$ GeV, a value of $m_\nu \sim 0.6$ eV is obtained.  This is just 
one order of magnitude greater than the square root of the $\Delta m^2 \sim 
5 \times 10^{-3}$ eV$^2$ needed for atmospheric neutrino oscillations 
\cite{1}.  Reducing slightly the above dimensionless couplings from unity 
would fit the data quite well.  Since $m_\chi \sim 10^{13}$ GeV is now 
allowed, leptogenesis should be possible as demonstrated in Ref.~[6].

\section{Neutrino Oscillations}

It has recently been shown \cite{15} that the structure of Eq.~(19) for the 
$\mu - \tau$ sector is naturally suited for the large mixing solution of 
atmospheric neutrino oscillations.  To be more specific, the $2 \times 2$ 
submatrix of Eq.~(19) for the $\mu - \tau$ sector can be written as
\begin{equation}
m_0 \left[ \begin{array} {c@{\quad}c} 2 \sin \alpha \sin \alpha' & \sin 
(\alpha + \alpha') \\ \sin (\alpha + \alpha') & 2 \cos \alpha \cos \alpha' 
\end{array} \right],
\end{equation}
where $\tan \alpha = f_\mu/f_\tau$ and $\tan \alpha' = f'_\mu/f'_\tau$. 
The eigenvalues of the above are then given by $m_0 (c_1 \pm 1)$, where 
$c_1 = \cos (\alpha - \alpha')$, and the effective $\sin^2 2 \theta$ for 
$\nu_\mu - \nu_\tau$ oscillations is $(1-c_2)/(1+c_2)$, where $c_2 = \cos 
(\alpha + \alpha')$.  If $\tan \alpha \sim \tan \alpha' \sim 1$, then 
$c_1 \sim 1$ and $c_2 \sim 0$.  In that case, maximal mixing between a 
heavy $(2 m_0)$ and a light $(s_1^2 m_0/2)$ neutrino occurs as an 
explanation of the atmospheric data.  If it is assumed further that 
$f_e << f_{\mu,\tau}$ and $f'_e << f'_{\mu,\tau}$, then the small-angle 
matter-enhanced solution of solar neutrino oscillations may be obtained 
as well.

\section{Collider Phenomenology}

The above model has the twin virtues of an acceptable neutrino mass matrix 
and the possibility of generating a lepton asymmetry of the Universe.  It is 
also phenomenologically safe because all the additions to the standard model 
do not alter its known successes.  Neither the fourth family of leptons nor 
the two extra Higgs doublets mix with their standard-model analogs because 
they are odd under the new discrete $Z'_2$ symmetry.  In particular, $H_3$ 
and $H_4$ do not couple to the known quarks and leptons, hence flavor-changing 
neutral currents are suppressed here as in the standard model.  The 
lepton-number violation of this model is associated with $m_\chi$ which is 
of order $10^{13}$ GeV.  However, the fourth family of leptons should have 
masses of order $10^2$ GeV and be observable at planned future colliders. 
The two extra Higgs doublets should also be observable with an energy scale 
of order $M_{SUSY}$.  The soft supersymmetry-breaking terms of this model 
are assumed to break $Z'_2$ without breaking $Z_2$.  Hence there will still 
be a stable LSP (lightest supersymmetric particle) and a fourth-family 
lepton will still decay into ordinary leptons.  For example, because $\tilde 
h_3^0$ mixes with $\tilde h_1^0$, the decay
\begin{equation}
E^- \to \mu^- \tilde h_3^0 (\tilde h_1^0) \to \mu^- \tau^+ \tau^-
\end{equation}
is possible and would make a spectacular signature.

\section{Conclusion}

In conclusion, the issue of neutrino masses in supersymmetry has been 
addressed in this talk.  The assumption of $R$-parity nonconservation is 
shown to be generically inconsistent with leptogenesis because the 
lepton-number violating interactions would act in conjunction with the $B+L$ 
violating sphaleron processes and erase any pre-existing $B$ or $L$ or 
$B-L$ asymmetry of the Universe.  This constraint means that any $R$-parity 
violation must be very small, so that it is of negligible phenomenological 
interest and cannot contribute significantly to neutrino masses.  This 
conclusion also applies to models of radiative neutrino masses with 
suppressed Yukawa couplings, such as the Zee model.  However, it has also 
been shown that realistic neutrino masses in supersymmetry are possible 
beyond the MSSM with $R$-parity conservation where lepton-number 
violation is by two units and occurs at the mass scale of $10^{13}$ GeV. 
The specific model presented also predicts new particles which should be 
observable in the future at the LHC (Large Hadron Collider).

\bigskip

\begin{center} {ACKNOWLEDGEMENT}
\end{center}

I thank the organizers George Zoupanos, Nick Tracas, and George Koutsoumbas 
for their great hospitality at Corfu.  This work was supported in part by the 
U.~S.~Department of Energy under Grant No.~DE-FG03-94ER40837.

\end{document}